# Vortex creation in Bose–Einstein condensates by diffraction on a helical light grating

Kestutis Staliunas

*Physikalisch Technische Bundesanstalt, Bundesallee 100, D-38116 Braunschweig, Germany.*

*kestutis.staliunas@ptb.de*

**Abstract**

It is shown that diffraction of Bose-Einstein condensates on a helical light grating results in vortices in the condensate. Helical light gratings can be produced by light interference of a Gaussian laser beam with a helical one. Diffraction orders of the Bose-Einstein condensate contain vortices of corresponding topological charge. The phenomena is illustrated by numerical integration of the Gross-Pitaevskii equation in two dimensions.

Various techniques for producing of vortices in Bose-Einstein condensates have been proposed. They are based on the rotation of magnetic traps [1,2], on rotational stirring of the condensate with a laser beam [3], on imprinting a desired phase distribution onto the condensate by optical means [4], or on vortex guiding by a laser beam [5]. Some of these suggestions lead to experimental demonstration of vortices in two component condensates [6], in a single component condensate [7], and to realisation of large ensembles of vortices [8]. Here we suggest alternatively, that vortices and vortex ensembles can be effectively created by the diffraction of a condensate on helical light gratings.

Diffraction of Bose-Einstein condensates on light gratings has been shown experimentally [9-11], where the diffraction grating was produced by interference of counterpropagating laser beams. The light pattern introduces a repulsive (attractive) potential for blue (red) detuned light, and can therefore be used to impose phase modulation of the condensate [12]. For example the light intensity pattern of lamella form causes a corresponding modulation of the phase of the condensate, and as a result the condensate splits into several parts - the central part (zero order), and two first order diffraction components moving along the normal of the light interference grating.

Here we suggest, that the light interference grating be produced by the interference between a Gaussian laser beam and a helical one, such as a Gauss-Laguerre (doughnut) mode TEM$^*_{01}$. The illustration is given in Fig.1: A Gaussian beam with the spatial field distribution: $E_0(\mathbf{r}) = A_0 \exp(ikz - (x^2 + y^2)/r_0^2)$ propagating along the z-axis interferes with a counterpropagating helical beam $E_1(\mathbf{r}) = A_1 \cdot (x+iy)\exp(-ikz - (x^2+y^2)/r_0^2)$. This results in intensity distribution of the interference pattern $I(\mathbf{r}) = |E_0(\mathbf{r}) + E_1(\mathbf{r})|^2$:

$$I(\mathbf{r}) = (|A_0|^2 + (x^2+y^2)|A_1|^2)\exp(-2(x^2+y^2)/r_0^2) +$$
$$A_0^* A_1 \cdot (x+iy)\exp(-2ikz - 2(x^2+y^2)/r_0^2) +$$
$$A_0 A_1^* \cdot (x-iy)\exp(2ikz - 2(x^2+y^2)/r_0^2) \qquad (1)$$

The light interference pattern then consists of a part modulated on the scale of the width of the laser beam $r_0$ (first term in (1)), as well as of the helical light grating (second and third terms of (1)). The helical light interference grating is shown in Fig.1 as given by the expression (1). The sign of helicity of the light grating is determined by the charge of the optical vortex of the helical light beam.

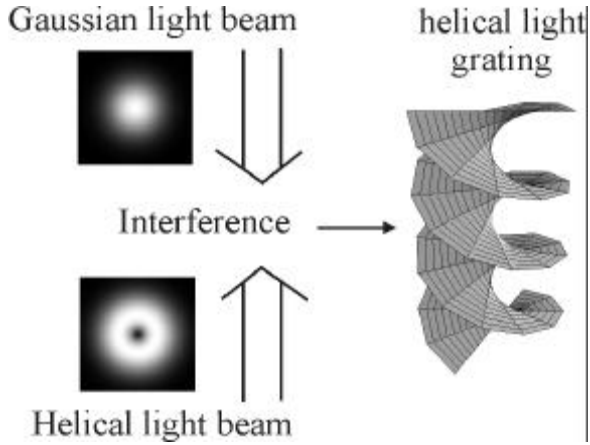

*Fig.1. The light intensity pattern resulting from the interference of a Gaussian laser beam (down-propagating) with a helical beam (up-propagating). Helical surface give the maximum light intensity.*

Next we consider the interaction of a Bose-Einstein condensate with the above helical light interference pattern. The interaction results in a modulation of the phase of the condensate proportional to the light intensity $I(\mathbf{r})$, and exposure time $\Delta t$: $y_1(\mathbf{r}) = y_0(\mathbf{r})\exp(ia I(\mathbf{r})\Delta t)$, where $a$ is the coefficient dependent on the light interaction with matter; $y_0(\mathbf{r})$ and $y_1(\mathbf{r})$ are the condensate wave-functions before and after the interaction with light. The subsequent evolution of the condensate can be calculated numerically, e.g. by solving numerically the corresponding three-dimensional Gross-Pitaevskii equation. However it is sufficient, in order to show the phenomena, not to integrate the nonlinear dynamical equation, but assume a linear evolution of the condensate with the initial distribution $y_1(\mathbf{r})$. The momentum distribution of the condensate is given by the three-dimensional Fourier transformation of the condensate wave-function $y_1(\mathbf{r})$. In case of weak phase modulation, the wave-function can be Taylor-approximated by: $y_1(\mathbf{r}) = y_0(\mathbf{r})\exp(iaI(\mathbf{r})) = y_0(\mathbf{r}) \cdot (1 + iaI(\mathbf{r}))$. For an initial Gaussian distribution of the condensate $y_0(\mathbf{r}) = \exp(-\mathbf{r}^2/r_c^2)$, and for an asymptotically broad laser beam $r_0 \gg r_c$ the Fourier transformation leads to a simple analytical expression:

$$y(\mathbf{k}) = y_0(\mathbf{k})(1 + ia\Delta t(|A_0|^2 + r_c^2|A_1|^2(1 - r_c^2(k_x^2 + k_y^2)/4)) -$$
$$y_0(\mathbf{k} + 2k_z\mathbf{e}_z)A_0^* A_1 \cdot (k_x + ik_y) -$$
$$y_0(\mathbf{k} - 2k_z\mathbf{e}_z)A_0 A_1^* \cdot (k_x - ik_y) \qquad (2)$$

here $\psi_0(\mathbf{k}) = \pi^{3/2} r_c^3 \exp(-\mathbf{k}^2 r_c^2 / 4)$ is the Fourier transformation of the unperturbed Bose-Einstein condensate $\psi_0(\mathbf{r})$.

The momentum distribution (2) is illustrated in Fig.2. It consists of three components: the central component (first term in (2)), and two first order diffraction components (second and third term in (2)). The first order diffraction components are shifted by $\pm 2k_z \mathbf{e}_z$ in the momentum space, and are of toroidal shapes, i.e. contain vortex lines, as shown in Fig.2. For sufficiently large values of $k_z$ the components of the condensate are well separated in momentum space.

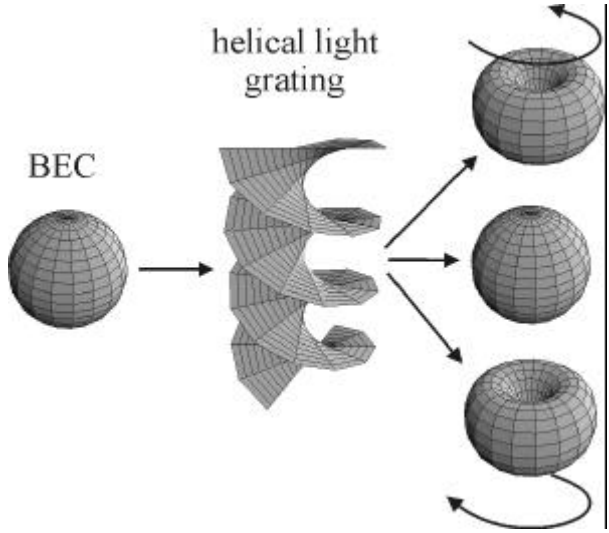

***Fig.2***. *Diffraction of a Bose-Einstein condensate on the helical light grating, as obtained from (2). Two first order diffraction components are created, propagating upwards (with +1 topological charge) and downwards (with +1 topological charge).*

If the trapping potential is removed immediately after imposing the phase modulation, then the evolution of the condensate is well described by the linearized version of Gross-Pitaevkii equation [13]:

$$i \frac{\partial \psi(\mathbf{r},t)}{\partial t} = -\nabla^2 \psi(\mathbf{r},t) \qquad (3)$$

because the nonlinear interaction term for the expanding condensate becomes negligible.

In the linear propagation of the condensate, as described by (3) the three condensate components evolve (diffract) independently from one another, and also, due to the different centering in momentum space, move with different velocities. The central component, according to (3) diffracts, but her mass center remains at rest. The first order diffraction components also diffract, but move apart with velocities $\mathbf{v}_{\pm 1} = \pm 4k_z \mathbf{e}_z$. Asymptotically, for $t \to \infty$, the distributions of the wavefunction in coordinate space are identical to the distributions in momentum space (2). I.e. the components can be well separated, and contain vortex lines, as illustrated in Fig.2.

An alternative technique is to retain the trapping potentials after imposing modulation of the phase. In this case the wave evolution of the wave-function is governed by the full Gross-Pitaevkii equation [13]:

$$i \frac{\partial \psi(\mathbf{r},t)}{\partial t} = -\nabla^2 \psi(\mathbf{r},t) + V(\mathbf{r},t)\psi(\mathbf{r},t) + C|\psi(\mathbf{r},t)|^2 \psi(\mathbf{r},t) \qquad (4)$$

Here the external potential corresponds to the harmonic magnetic trap $V(\mathbf{r}) = r^2/4$, the time is in units of the inverse frequency of the harmonic trap $\omega_t^{-1}$, and the spatial coordinates are in units of the size of the noninteracting condensate: $(\hbar/(2m\omega_t))^{1/2}$, where $m$ is the mass of the atoms. The density of the wave-function of the condensate $\rho(\mathbf{r},t) = |\psi(\mathbf{r},t)|^2$ is normalized: $\int \rho(\mathbf{r},t)d\mathbf{r} = 1$, and the coefficient $C$ is proportional to the number of atoms in the trap $N$ and their scattering length $a$: $C = 2Na(2m\omega_t/\hbar)^{1/2}$. For simplicity a two-dimensional condensate (of a "pancake" shape) was considered.

The results of the numerical integration are given in Fig.3, where the coefficient was fixed to $C = 180$, which corresponds to e.g. $N = 10^5$ of $^{87}Rb$ atoms ($a = 5.2*10^{-9}$ m), in a magnetic trap of frequency $\omega_t = 20 \times \pi$ rad s$^{-1}$, and in a spatial size (diameter at half of the maximum density) of the condensate of 50 $\mu$m.

The stationary condensate was subjected to the phase modulation, where the form or the phase mask is a stripe structure with a dislocation (see Fig.3. at time $t = 0$). The depth of the phase modulation was chosen equal to $2\pi$, which allows to eliminate the central component of the diffracted condensate. Experimentally such phase mask can be formed by visible laser beams ($\lambda = 0.5$ $\mu$m) interfering at an angle of approximately 3 degrees (or alternatively by suitable holograms [14]. After the phase mask is switched off the condensate splits into two components with opposite momenta. The components of the condensate separate maximally after 1/4 of the trap period. The vortices are most symmetric at this time of maximal separation. However, due to the trap potential, the condensates fall back into the middle of the trap, which results in the centrosymetrically inverted picture of the initial condensate. At $t = 3/4$ the condensate splits maximally again, however the charges of vortices are opposite to those at t=1/4. This scenario of periodic revivals continues, despite the nonlinearity in the Gross-Pitaevskii equation.

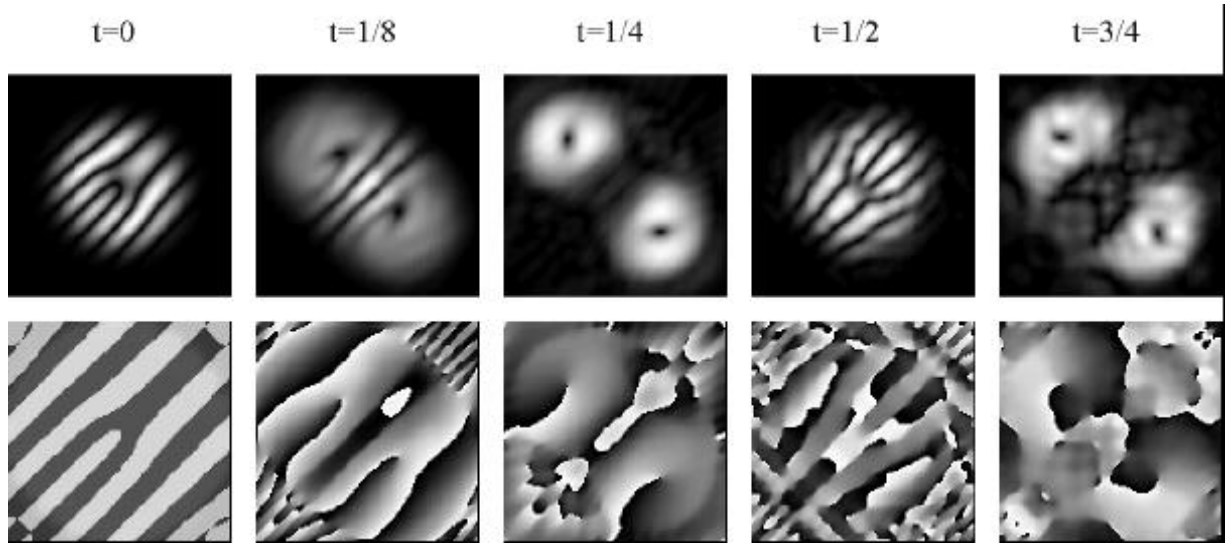

*Fig.3. Diffraction of two-dimensional (pancake), trapped Bose-Einstein condensate on light grating, as obtained from numerical integration of (4). Upper row - density, lower row - phase of the condensate. At t=0 the grating was imposed, imprinting the corresponding modulation of the density and phase of the condensate; at t=1/8 (time is in the units of the trap period) the vortices emerge; at t/4 the splitting of condensate is maximal. The vortices are of opposite charge, as visible from the phase picture; at t=1/2 the condensate falls back into the center of the trap; at t=3/4 the condensate splits maximally again, however the charges of the vortices are opposite to those at t=1/4.*

From the above study we draw the following conclusions:

1) The first order Taylor approximation yields only two first order diffraction components. In general higher order diffraction maxima should be present. The second order components contain then vortices of +2 and -2 topological charge, as an example.

2) If the laser beams are not strictly counterpropagating but interfere at some angle the vortices are nonetheless generated in the diffracted Bose-Einstein condensates. The vortex lines are then directed along the direction of the helical light beam propagation.

3) The vortices can be observation in the ballistic expansion phase of the coherent atom ensembles (when the trapping potential is removed immediately after imposing phase modulation) as well as in the trap. A specially designed (nonstationary) traps could allow to separate the components of condensate with vortices.

4) The technique can be applied also for Bragg diffraction on a „thick" grating. Bragg diffraction has been demonstrated for trapped condensates as well as for coherent atom ensembles, using moving diffraction gratings [9]. A moving helical diffraction grating, created by interference of Gaussian and helical light beams of different frequency, would also result in Bragg diffraction. Bragg diffraction should produce only one first order component.

5) The technique can be generalized to more complicated light gratings (holograms), which would allow to produce of complicated vortex ensembles.

**Acknowledgement:**

The work was supported by Sonderforschungsbereich 407 of Deutsche Forschungsgemeinschaft. Discussions with C.O.Weiss, M.Lewenstein, and L.Santos are gratefully acknowledged.